\documentstyle[aps,prl,epsf,amssymb,amsmath,amsthm,eqsecnum]{revtex}
\bibstyle{unsrt}

\tighten
\begin{document}
\draft \preprint{}


\title{Calculation of the Hidden Symmetry Operator in
${\cal PT}$-Symmetric Quantum Mechanics}

\author{Carl M. Bender, Peter N. Meisinger, and Qinghai Wang}

\address{Department of Physics, Washington University, St. Louis, MO 63130, USA}

\date{\today}
\maketitle

\begin{abstract}
In a recent paper it was shown that if a Hamiltonian $H$ has an unbroken ${\cal
PT}$ symmetry, then it also possesses a hidden symmetry represented by the
linear operator ${\cal C}$. The operator ${\cal C}$ commutes with both $H$ and
${\cal PT}$. The inner product with respect to ${\cal CPT}$ is associated with a
positive norm and the quantum theory built on the associated Hilbert space is
unitary. In this paper it is shown how to construct the operator ${\cal C}$ for
the non-Hermitian ${\cal PT}$-symmetric Hamiltonian $H={1\over2}p^2+{1\over2}x^2
+i\epsilon x^3$ using perturbative techniques. It is also shown how to construct
the operator ${\cal C}$ for $H={1\over2}p^2+{1\over2}x^2-\epsilon x^4$ using
nonperturbative methods.
\end{abstract}

\vskip2pc

\section{Introduction and Background}
\label{s1}

It was observed in 1998 \cite{BB} that with properly defined boundary conditions
the Sturm-Liouville differential equation eigenvalue problem associated with the
non-Hermitian Hamiltonian
\begin{equation}
H=p^2+x^2(ix)^\nu\qquad(\nu>0)
\label{e1}
\end{equation}
exhibits a spectrum that is {\em real and positive}. It was argued in
Ref.~\cite{BB} that the reality of the spectrum of $H$ is a consequence of the
unbroken ${\cal PT}$ symmetry of $H$. A complete proof that the spectrum of $H$
is real and positive was given by Dorey et {\em al} \cite{DDT}.

By ${\cal PT}$ symmetry we mean the following: The linear parity operator ${\cal
P}$ performs spatial reflection and thus reverses the sign of the momentum and
position operators: ${\cal P}p{\cal P}^{-1}=-p$ and ${\cal P}x{\cal P}^{-1}=-x$.
The antilinear time-reversal operator ${\cal T}$ reverses the sign of the
momentum operator and performs complex conjugation: ${\cal T}p{\cal T}^{-1}=-p$,
${\cal T}x{\cal T}^{-1}=x$, and ${\cal T}i{\cal T}^{-1}=-i$. The Heisenberg
algebra, $[x,p]=i$, which is fundamental in quantum theory because it embodies
the uncertainty principle, is invariant under the action of the operators ${\cal
P}$ and ${\cal T}$ separately. The non-Hermitian Hamiltonian $H$ in (\ref{e1})
is not symmetric under ${\cal P}$ or ${\cal T}$ separately, but it {\em is}
invariant under their combined operation; such Hamiltonians are said to possess
{\em space-time reflection symmetry} (${\cal PT}$ symmetry).

We say that the ${\cal PT}$ symmetry of a Hamiltonian $H$ is {\em not
spontaneously broken} if the eigenfunctions of $H$ are simultaneously
eigenfunctions of the ${\cal PT}$ operator. It is difficult to prove that the
${\cal PT}$ symmetry of a given Hamiltonian is not spontaneously broken, but if
this is the case, then it is easy to show that the spectrum is entirely real
\cite{BBM}.

Space-time reflection (${\cal PT}$) symmetry is a weaker condition than
Hermiticity in the following sense. For many different Hermitian Hamiltonians,
such as $H=p^2+x^4$, $H=p^2+x^6$, $H=p^2+x^8$, and so on, we can construct
infinite classes of non-Hermitian ${\cal PT}$-symmetric Hamiltonians $H=p^2+x^4
(ix)^\nu$, $H=p^2+x^6(ix)^\nu$, $H=p^2+x^8(ix)^\nu$, and so on. So long as the
parameter $\nu$ is real and positive ($\nu>0$), the ${\cal PT}$ symmetry of each
of these Hamiltonians is not spontaneously broken and the spectrum is entirely
real \cite{BBM}.

Showing that the Sturm-Liouville problem associated with a non-Hermitian ${\cal
PT}$-symmetric Hamiltonian has a positive real spectrum is mathematically
significant, but it does not have any obvious relevance to physics. To show that
a Hamiltonian can serve as the basis for a theory of quantum mechanics it is
necessary to demonstrate that the Hamiltonian acts on a Hilbert space that is
endowed with an inner product whose associated norm is positive definite. Only
then can one say that the theory has a probabilistic interpretation.
Furthermore, it must be shown that the theory is unitary (probability must be
conserved in time). Since the publication of Ref.~\cite{BB} it has been believed
that the Hamiltonians in (\ref{e1}) could not be the basis for a physical theory
because they are non-Hermitian. Indeed, the ${\cal PT}$ norm is not positive
definite and this appears to present interpretational problems in developing a
quantum theory based on ${\cal PT}$-symmetric Hamiltonians. Many papers have
been published that discuss this apparent shortcoming of non-Hermitian ${\cal
PT}$-symmetric Hamiltonians \cite{MAZM}.

In a recent letter it was shown how to overcome these problems \cite{BBJ}. This
letter demonstrates that any Hamiltonian that possesses an unbroken ${\cal PT}$
symmetry also has a hidden symmetry. This new symmetry is represented by the
linear operator ${\cal C}$, which commutes with both the Hamiltonian $H$ and the
${\cal PT}$ operator. In terms of ${\cal C}$ one can construct an inner product
whose associated norm is positive definite. Observables exhibit ${\cal CPT}$
symmetry and the dynamics is governed by unitary time evolution. Thus, ${\cal
PT}$-symmetric Hamiltonians give rise to new classes of fully consistent complex
quantum theories. These new quantum theories are extensions of conventional
Hermitian quantum mechanics into the complex domain. The novelty of these
theories is that the inner product is not specified prior to and independently
of the Hamiltonian. Rather, the inner product is {\em determined} by the
Hamiltonian itself. Thus, in such theories the norm and hence the notion of
probability is dynamically incorporated in the Hamiltonian.

The purpose of the present paper is to present an explicit calculation of ${\cal
C}$ for two nontrivial Hamiltonians. First, we consider the case of the ${\cal
PT}$-symmetric Hamiltonian
\begin{equation}
H={1\over2}p^2+{1\over2}x^2+i\epsilon x^3,
\label{e2}
\end{equation}
for which we give a perturbative calculation of the operator ${\cal C}$ correct
to {\em third} order in powers of $\epsilon$. Second, we calculate ${\cal C}$
for the Hamiltonian
\begin{equation}
H={1\over2}p^2+{1\over2}x^2-\epsilon x^4,
\label{e3}
\end{equation}
for which ordinary perturbative methods are ineffective and nonperturbative
methods must be used. The organization of this paper is straightforward. In
Sec.~\ref{s2} we review the formal construction, first presented in
Ref.~\cite{BBJ}, of the ${\cal C}$ operator. In Sec.~\ref{s3} we calculate
${\cal C}$ for the Hamiltonian in (\ref{e2}) and in Sec.~\ref{s4} we calculate
${\cal C}$ for the Hamiltonian in (\ref{e3}).

\section{Formal Derivation of the ${\cal C}$ Operator}
\label{s2}

In this section we present a formal discussion of ${\cal PT}$-symmetric
Hamiltonians and we show how to construct the ${\cal C}$ operator. In general,
for any ${\cal PT}$-symmetric Hamiltonian $H$ we must begin by solving the
Sturm-Liouville differential equation eigenvalue problem associated with $H$:
\begin{equation}
H\phi_n(x)=E_n\phi_n(x)\qquad(n=0,\,1,\,2,\,3,\,\cdots).
\label{e4}
\end{equation}

For Hamiltonians like those in (\ref{e1} -- \ref{e3}) the differential equation
(\ref{e4}) must be imposed on an infinite contour in the complex-$x$ plane. For
large $|x|$ the contour lies in wedges that are placed symmetrically with
respect to the imaginary-$x$ axis. These wedges are described in Ref.~\cite{BB}.
The boundary conditions on the eigenfunctions are that $\phi(x)\to0$
exponentially rapidly as $|x|\to\infty$ on the contour. For $H$ in (\ref{e2})
the contour may be taken to be the real-$x$ axis, but for $H$ in (\ref{e3}) the
contour lies in the two wedges $-\pi/3<{\rm arg}\,x<0$ and $-\pi<{\rm arg}\,x<-2
\pi/3$. It is not possible to solve the differential equation (\ref{e4})
analytically for the two Hamiltonians (\ref{e2}) and (\ref{e3}) considered in
this paper but we have solved it numerically to very high accuracy for the first
ten eigenfunctions and eigenvalues. As mentioned above, the eigenvalues are all
real and positive and are nondegenerate.

For all $n$, the eigenfunctions $\phi_n(x)$ are simultaneously eigenstates of
the ${\cal PT}$ operator: ${\cal PT}\phi_n(x)=\lambda_n\phi_n(x)$. Moreover,
because $({\cal PT})^2=1$ and ${\cal PT}$ involves complex conjugation, it
follows that $|\lambda_n|=1$. Thus, $\lambda_n=e^{i\alpha_n}$ is a pure phase.
For each $n$ this phase can be absorbed into $\phi_n$ by the multiplicative
rescaling $\phi_n\to e^{-i\alpha_n/2}\phi_n$, so that the new eigenvalue of
${\cal PT}$ is unity:
\begin{equation}
{\cal PT}\phi_n(x)=\phi_n(x).
\label{e5}
\end{equation}

Next, we observe that there is an inner product, called the ${\cal PT}$ inner
product, with respect to which the eigenfunctions $\phi_n(x)$ for two different
values of $n$ are orthogonal. For the two functions $f(x)$ and $g(x)$ the ${\cal
PT}$ inner product $(f,g)$ is defined by
\begin{equation}
(f,g)\equiv\int_{\rm{C}} dx\,\left[{\cal PT}f(x)\right]g(x),
\label{e6}
\end{equation}
where ${\cal PT}f(x)=\left[f(-x)\right]^*$ and the contour ${\rm C}$ lies in
the wedges described above. For this inner product the associated norm $(f,f)$
is independent of the overall phase of $f(x)$ and is conserved in time. (Phase
independence is required because ultimately we must construct a space of rays to
represent quantum mechanical states.) The proof that eigenfunctions $\phi_n(x)$
corresponding to different values of $n$ are orthogonal with respect to this
inner product is trivial and follows directly from the differential equation
(\ref{e4}) using integration by parts.

We then normalize the eigenfunctions so that $|(\phi_n,\phi_n)|=1$ and we
discover the apparent problem with using a non-Hermitian Hamiltonian. While the
eigenfunctions are orthogonal, the ${\cal PT}$ norm is not positive definite:
\begin{equation}
(\phi_m,\phi_n)=(-1)^n \delta_{m,n} \qquad(m,\,n=0,\,1,\,2,\,3,\,\cdots).
\label{e7}
\end{equation}

Despite the fact that this norm is not positive definite, the eigenfunctions are
complete. For real $x$ and $y$ the statement of completeness in coordinate space
is\footnote{It is important to remark here that the argument of the Dirac delta
function in (\ref{e8}) must be {\em real} because the delta function is only
defined for real argument. This may seem to be in conflict with the earlier
remark in this section that the Schr\"odinger equation (\ref{e4}) must be solved
along a contour that lies in wedges in the complex-$x$ plane. To resolve this
apparent conflict we specify the contour as follows. We demand that the contour
lie on the real axis until it passes the points $x$ and $y$. Only then may it
veer off into the complex-$x$ plane and enter the wedges. This choice of contour
is allowed because the wedge conditions are {\em asymptotic} conditions. The
positions of the wedges are determined by the boundary conditions.}
\begin{equation}
\sum_n(-1)^n \phi_n(x)\phi_n(y)=\delta(x-y).
\label{e8}
\end{equation}
This is a nontrivial result that has been verified numerically to extremely high
accuracy \cite{BW}. Using (\ref{e7}) we can verify that the sum in
(\ref{e8}) is the position-space representation of the unity operator:
\begin{equation}
\int dy\,\delta(x-y)\delta(y-z)=\delta(x-z).
\label{e9}
\end{equation}

We can also express the Hamiltonian $H$ and the Green's function $G(x,y)$ in
the coordinate space representation:
\begin{equation}
H(x,y)=\sum_n(-1)^n E_n\phi_n(x)\phi_n(y)\qquad{\rm and}\qquad
G(x,y)=\sum_n(-1)^n{1\over E_n}\phi_n(x)\phi_n(y).
\label{e11}
\end{equation}
The Green's function $G(x,y)$ satisfies the inhomogeneous differential equation
\begin{equation}
H G(x,y)=\delta(x-y),
\label{e12}
\end{equation}
which states that the Green's function is the inverse of the Hamiltonian
operator.

In addition, we can construct the parity operator ${\cal P}$ in terms of the
energy eigenstates. In position space
\begin{equation}
{\cal P}(x,y)=\delta(x+y)=\sum_n(-1)^n\phi_n(x)\phi_n(-y).
\label{e13}
\end{equation}
Again, using (\ref{e7}) we can see that the square of the parity operator is
unity.

Finally, we construct the linear operator ${\cal C}$ that expresses the hidden
symmetry of the Hamiltonian $H$. The position-space representation of ${\cal C}$
is
\begin{equation}
{\cal C}(x,y)=\sum_n\phi_n(x)\phi_n(y).
\label{e14}
\end{equation}
The properties of the operator ${\cal C}$ are easy to verify using (\ref{e7}).
First, like the parity operator, the square of ${\cal C}$ is unity:
\begin{equation}
\int dy\,{\cal C}(x,y){\cal C}(y,z)=\delta(x-z).
\label{e15}
\end{equation}
Second, the eigenfunctions $\phi_n(x)$ of the Hamiltonian $H$ are also
eigenfunctions of ${\cal C}$ and the corresponding eigenvalues are $(-1)^n$:
\begin{equation}
\int dy\,{\cal C}(x,y)\phi_n(y)=(-1)^n\phi_n(x).
\label{e16}
\end{equation}
Third, the operator ${\cal C}$ commutes with both the Hamiltonian $H$ and the
operator ${\cal PT}$. Note that while the operators ${\cal P}$ and ${\cal C}$
are unequal (the parity operator ${\cal P}$ is real, while the operator ${\cal
C}$ is complex), both ${\cal P}$ and ${\cal C}$ are square roots of the unity
operator $\delta(x-y)$. Last, the operators ${\cal P}$ and ${\cal C}$ do not
commute. Indeed, ${\cal CP}=({\cal PC})^*$.

The operator ${\cal C}$ does not exist as a distinct entity in conventional
Hermitian quantum mechanics. Indeed, we will see that as the parameter
$\epsilon$ in (\ref{e2}) and (\ref{e3}) tends to zero the operator ${\cal C}$
becomes identical to ${\cal P}$. Thus, in this limit the ${\cal CPT}$ operator
becomes ${\cal T}$. This verifies that for symmetric Hamiltonians in standard
quantum mechanics ${\cal CPT}$ symmetry and Hermiticity coincide and ${\cal
CPT}$ invariance can be viewed as the natural complex extension of the usual
Hermiticity condition.

We can now define an inner product $\langle f|g\rangle$ whose associated norm
is positive:
\begin{equation}
\langle f|g\rangle\equiv \int dx\,[{\cal CPT}f(x)]g(x).
\label{e17}
\end{equation}
The ${\cal CPT}$ norm associated with this inner product is positive because
${\cal C}$ contributes $-1$ when it acts on states with negative ${\cal PT}$
norm. To verify that this norm is positive definite we expand an arbitrary
function $f(x)$ as a linear combination of eigenfunctions of the Hamiltonian
$H$:
$$f(x)=\sum_{n=0}^\infty c_n \phi_n(x).$$
Then, the ${\cal CPT}$ norm of $f(x)$ is
$$\langle f|f\rangle=\int_{-\infty}^\infty dx\,[{\cal
CPT}f(x)]f(x)=\sum_{n=0}^\infty |c_n|^2,$$
which is positive unless $f(x)\equiv0$. The ${\cal CPT}$ norm is time
independent because the ${\cal CPT}$ operator commutes with the Hamiltonian $H$
and thus the theory is unitary. Using the ${\cal CPT}$ conjugate, the
completeness condition (\ref{e8}) becomes
\begin{equation}
\sum_n [{\cal CPT}\phi_n(x)]\phi_n(y)=\delta(x-y).
\label{e18}
\end{equation}

\section{Perturbative Calculation of ${\cal C}$ in a ${\cal PT}$-Symmetric Cubic
Theory}
\label{s3}

In this section we use perturbative methods to calculate the operator ${\cal C}
(x,y)$ for the Hamiltonian $H={1\over2}p^2+{1\over2}x^2+i\epsilon x^3$. We
perform the calculations to third order in perturbation theory. We begin by
solving the Schr\"odinger equation
\begin{equation}
-{1\over2}\phi_n''(x)+{1\over2}x^2\phi_n(x)+i\epsilon x^3\phi_n(x)=E_n\phi_n(x)
\label{e31}
\end{equation}
as a series in powers of $\epsilon$.

The perturbative solution to this equation has the form
\begin{equation}
\phi_n(x)={i^n a_n\over\pi^{1/4} 2^{n/2}\sqrt{n!}} e^{-{1\over2}x^2}
\left[H_n(x)-iP_n(x)\epsilon -Q_n(x)\epsilon^2+iR_n(x)\epsilon^3\right],
\label{e32}
\end{equation}
where $H_n(x)$ is the $n$th Hermite polynomial and $P_n(x)$, $Q_n(x)$, and
$R_n(x)$ are polynomials in $x$ of degree $n+3$, $n+6$, and $n+9$,
respectively. These polynomials can be expressed as linear combinations of
Hermite polynomials:

\begin{eqnarray}
P_n(x) &=& {1\over24}H_{n+3}(x)+{3\over4}(n+1)H_{n+1}(x)
-{3\over2}n^2 H_{n-1}(x)-{1\over3}n(n-1)(n-2) H_{n-3}(x),\nonumber\\
Q_n(x) &=& {1\over1152}H_{n+6}(x)+{1\over128}(4n+7)H_{n+4}(x)
+{1\over32}(7n^2+33n+27) H_{n+2}(x)\nonumber\\
&&\qquad +{1\over8}n(n-1)(7n^2-19n+1)H_{n-2}(x)+{1\over8}n(n-1)(n-2)(n-3)(4n-3)
H_{n-4}(x) \nonumber\\
&&\qquad +{1\over18}n(n-1)(n-2)(n-3)(n-4)(n-5) H_{n-6}(x),\nonumber\\
R_n(x) &=& {1\over82944}H_{n+9}(x)+{1\over3072}(2n+5)H_{n+7}(x)
+{1\over7680}(80n^2+465n+549) H_{n+5}(x)\nonumber\\
&&\qquad +{1\over6912}(488n^3+3639n^2+9832n+7506)H_{n+3}(x)\nonumber\\
&&\qquad +{3\over128}(20n^4-n^3+203n^2+408n+228)H_{n+1}(x) \nonumber\\
&&\qquad -{3\over64}n(20n^4+81n^3+326n^2+81n+44)H_{n-1}(x)\nonumber\\
&&\qquad -{1\over864}n(n-1)(n-2)(488n^3-2175n^2+4018n-825)H_{n-3}(x)\nonumber\\
&&\qquad -{1\over240}n(n-1)(n-2)(n-3)(n-4)(80n^2-305n+164)H_{n-5}(x)\nonumber\\
&&\qquad -{1\over24}n(n-1)(n-2)(n-3)(n-4)(n-5)(n-6)(2n-3)H_{n-7}(x)\nonumber\\
&&\qquad -{1\over162}n(n-1)(n-2)(n-3)(n-4)(n-5)(n-6)(n-7)(n-8)H_{n-9}(x).
\label{e33}
\end{eqnarray}
The energy $E_n$ to order $\epsilon^3$ is
\begin{equation}
E_n=n+{1\over2}+{1\over8}(30 n^2+30n+11)\epsilon^2+{\rm O}(\epsilon^4).
\label{e34}
\end{equation}
The expression for $\phi_n(x)$ must be ${\cal PT}$-normalized according to
(\ref{e7}) so that its square integral is $(-1)^n$:
\begin{equation}
\int_{-\infty}^{\infty} dx\,\left[\phi_n(x)\right]^2=(-1)^n+{\rm O}(\epsilon^4).
\label{e35}
\end{equation}
This determines the value of $a_n$ in (\ref{e32}):
\begin{equation}
a_n=1+{1\over144}(2n+1)(82n^2+82n+87)\epsilon^2+{\rm O}(\epsilon^4).
\label{e36}
\end{equation}

We calculate the operator ${\cal C}(x,y)=\sum_{n=0}^\infty\phi_n(x)\phi_n(y)$,
which is given formally in (\ref{e14}), by directly substituting the wave
functions $\phi_n(x)$ in (\ref{e32}). We then use the completeness relation for
Hermite polynomials,
\begin{equation}
{1\over\sqrt{\pi}}e^{-{1\over2}(x^2+y^2)}\sum_{n=0}^\infty{1\over2^nn!}H_n(x)
H_n(y)=\delta(x-y),
\label{e37}
\end{equation}
to evaluate the sum. We also need to use the following identities satisfied by
the Hermite polynomials:
\begin{eqnarray}
x H_n(x) &=& {1\over2}H_{n+1}(x)+nH_{n-1}(x),\nonumber\\
H_n''(x) &=& 2 x H_n'(x)-2nH_n(x),\nonumber\\
H_n'(x) &=& 2nH_{n-1}(x).
\label{e38}
\end{eqnarray}

To third order in $\epsilon$ the result is
\begin{eqnarray}
{\cal C}(x,y) &=&\left\{ 1-i\epsilon\left({4\over3}{\partial^3\over\partial x^3}
+2xy{\partial\over\partial x}\right)-\epsilon^2\left[{8\over9}{\partial^6\over
\partial x^6}+{8\over3}xy{\partial^4\over\partial x^4}+(2x^2y^2-12){\partial^2
\over\partial x^2}\right]+i\epsilon^3\left[{32\over81}{\partial^9\over\partial
x^9}+{16\over9}xy{\partial^7\over\partial x^7}\right.\right.\nonumber\\
&&\qquad \left.\left. +\left({8\over3}x^2y^2-{176\over5}
\right){\partial^5\over\partial x^5}+\left({4\over3}x^3y^3-48xy\right)
{\partial^3\over\partial x^3}+(-8x^2y^2+64)
{\partial\over\partial x}\right]+{\rm O}(\epsilon^4)\right\}\delta(x+y).
\label{e39}
\end{eqnarray}
Hence, the coordinate-space representation of the operator ${\cal C}(x,y)$ is
expressed as a derivative of a Dirac delta function. From this expression for
${\cal C}(x,y)$ we can verify the following properties: First, to order
$\epsilon^3$ the operator ${\cal C}(x,y)$ satisfies (\ref{e15}). That is,
\begin{equation}
\int_{-\infty}^\infty dy\,{\cal C}(x,y){\cal C}(y,z)=\delta(x-z)+{\rm O}
(\epsilon^4).
\label{e40}
\end{equation}
Second, to order $\epsilon^3$ the operator ${\cal C}(x,y)$ satisfies
(\ref{e16}); the wave functions $\phi_n(x)$ are eigenstates of ${\cal C}(x,y)$
with eigenvalue $(-1)^n$. That is,
\begin{equation}
\int_{-\infty}^\infty dy\,{\cal C}(x,y)\phi_n(y)=(-1)^n \phi_n(x)+{\rm O}
(\epsilon^4).
\label{e41}
\end{equation}
Third, in the limit as $\epsilon\to0$, the operator ${\cal C}(x,y)$ becomes the
coordinate-space representation of the parity operator ${\cal P}(x,y)=\delta
(x+y)$.

There is a somewhat simpler way to express the operator ${\cal C}(x,y)$. The
derivative operator in (\ref{e39}) that is acting on $\delta(x+y)$ can be
exponentiated so that to order $\epsilon^4$ (and not just $\epsilon^3$) we have
\begin{equation}
{\cal C}(x,y)=e^{-i\epsilon A-i\epsilon^3 B}\delta(x+y)+{\rm O}(\epsilon^5),
\label{e42}
\end{equation}
where the derivative operators $A$ and $B$ are given by
\begin{eqnarray}
A&=&{4\over3}{\partial^3\over\partial x^3}-2x{\partial\over\partial x}x
\nonumber\\
B&=&{128\over15}{\partial^5\over\partial x^5}-{40\over3}x
{\partial^3\over\partial x^3}x +8x^2 {\partial\over\partial x}x^2
-32 {\partial\over\partial x}.
\label{e43}
\end{eqnarray}

We have applied the procedure used above to calculate ${\cal C}(x,y)$ to
evaluate the parity operator ${\cal P}(x,y)$. That is, we have substituted the
eigenfunctions $\phi_n(x)$ in (\ref{e32}) into the formal sum in (\ref{e13}). We
find that to each order in powers of $\epsilon$ the summation {\em vanishes}
except for the leading term (the coefficient of $\epsilon^0$). Thus, we obtain
the result that ${\cal P}(x,y)=\delta(x+y)+{\rm O}(\epsilon^4)$. This is not a
new result, but it provides a useful check of the accuracy of our calculations.
Similarly, we have evaluated the sum in (\ref{e8}) and we obtain the trivial
result $\delta(x-y)+{\rm O}(\epsilon^4)$. We have also evaluated the expression
in (\ref{e11}) for the Hamiltonian in coordinate space and we find (as expected)
that the coefficient of $\epsilon^k$ in the summation vanishes for $k>1$ and
we get
$$H(x,y)=\left( -{1\over2}{\partial^2\over\partial x^2}+{1\over2}x^2+i\epsilon
x^3\right)\delta(x-y)+{\rm O}(\epsilon^4).$$

We have again applied the procedure for calculating ${\cal C}(x,y)$ to evaluate
the Green's function $G(x,y)$ in (\ref{e11}). Substituting the eigenfunctions
$\phi_n(x)$ in (\ref{e32}) into (\ref{e11}) and performing the summation gives
the perturbative expansion of the Green's function:
\begin{equation}
G(x,y)=G_0(x,y)-iG_1(x,y)\epsilon-G_2(x,y)\epsilon^2+iG_3(x,y)\epsilon^3+{\rm O}
(\epsilon^4).
\label{e44}
\end{equation}
The zeroth-order Green's function satisfies the differential equation
\begin{equation}
\left(-{1\over2}{\partial^2\over\partial x^2}+{1\over2}x^2\right)G_0(x,y)=\delta
(x-y).
\label{e45}
\end{equation}
The solution to this equation is
\begin{equation}
G_0(x,y)=\theta(x-y) D_{-1/2}(x\sqrt{2})D_{-1/2}(-y\sqrt{2})+\theta(y-x)D_{-1/2}
(-x\sqrt{2})D_{-1/2}(y\sqrt{2}),
\label{e46}
\end{equation}
where $D_\nu(x)$ is the parabolic cylinder function and $\theta(x)$ is the
step function defined by
\begin{equation}
\theta(x)=\left\{
\begin{array}{l}
0\quad (x<0),\cr
{1\over2}\quad (x=0),\cr
1\quad (x>0).
\end{array}
\right.
\label{e47}
\end{equation}
Note that $G_0(x,y)$ is a symmetric function of $x$ and $y$.

The first-order contribution to the Green's function satisfies the differential
equation
\begin{equation}
\left(-{1\over2}{\partial^2\over\partial x^2}+{1\over2}x^2\right)G_1(x,y)
=x^3 G_0(x,y)
\label{e48}
\end{equation}
and the solution to this equation is
\begin{equation}
G_1(x,y)=-{1\over3}\left( x^2{\partial\over\partial x}-x
+y^2{\partial\over\partial y}-y\right) G_0(x,y).
\label{e49}
\end{equation}
The second-order contribution to the Green's function satisfies
\begin{equation}
\left(-{1\over2}{\partial^2\over\partial x^2}+{1\over2}x^2\right)G_2(x,y)
=x^3 G_1(x,y)
\label{e50}
\end{equation}
and the solution to this equation is
\begin{equation}
G_2(x,y)={1\over18}\left( x^2{\partial\over\partial x}-x
+y^2{\partial\over\partial y}-y\right)^2 G_0(x,y)+{7\over6}\int_{-\infty}^\infty
dz\,z^4G_0(z,x)G_0(z,y).
\label{e51}
\end{equation}
The third-order contribution to the Green's function satisfies
\begin{equation}
\left(-{1\over2}{\partial^2\over\partial x^2}+{1\over2}x^2\right)G_3(x,y)
=x^3 G_2(x,y)
\label{e52}
\end{equation}
and the solution to this equation is
\begin{eqnarray}
G_3(x,y)&=&-{1\over9}\left[\left({5\over36}x^8+{1\over12}x^2y^6+{56\over15}x^4
+{112\over5}\right){\partial\over\partial x}+{25\over36}x^7-{1\over12}x^6y
-{112\over15}x^3\right.\nonumber\\
&&\qquad +\left.\left({5\over36}y^8+{1\over12}x^6y^2+{56\over15}y^4
+{112\over5}\right){\partial\over\partial y}+{25\over36}y^7-{1\over12}xy^6
-{112\over15}y^3\right]G_0(x,y)\nonumber\\
&&\qquad-{7\over12}\left(x^2{\partial\over\partial x}-x+y^2{\partial\over
\partial y}-y\right)\int_{-\infty}^\infty dz\,z^4G_0(z,x)G_0(z,y).
\label{e53}
\end{eqnarray}

\section{Nonperturbative Calculation of ${\cal C}$ in a ${\cal PT}$-Symmetric
Quartic Theory}
\label{s4}

In this section we explain briefly the nonperturbative methods that must be used
to calculate the operator ${\cal C}(x,y)$ for the Hamiltonian $H={1\over2}p^2+
{1\over2}x^2-\epsilon x^4$. We follow the approach taken in Ref.~\cite{BMY},
in which nonperturbative methods were used to calculate the one-point Green's
function for this Hamiltonian.

\subsection{Failure of Perturbation Theory}
\label{ss4.1}

We begin by explaining why perturbation theory fails to produce the operator
${\cal C}(x,y)$. Following the approach taken in Sec.~\ref{s3}, we expand the
solution to the Schr\"odinger equation
\begin{equation}
-{1\over2}\phi_n''(x)+{1\over2}x^2\phi_n(x)-\epsilon x^4\phi_n(x)=E_n\phi_n(x)
\label{e401}
\end{equation}
as a series in powers of $\epsilon$:
\begin{equation}
\phi_n(x)={i^n a_n\over\pi^{1/4}2^{n/2}\sqrt{n!}} e^{-{1\over2}x^2}
\left[H_n(x) +P_n(x)\epsilon\right] +{\rm O}(\epsilon^2),
\label{e402}
\end{equation}
where $H_n(x)$ is the $n$th Hermite polynomial and $P_n(x)$ is a polynomial in
$x$ of degree $n+4$. The polynomial $P_n(x)$ is a linear combination of Hermite
polynomials:
\begin{equation}
P_n(x)={1\over64}H_{n+4}(x)+{1\over8}(2n+3)H_{n+2}(x)
-{1\over2}n(n-1)(2n-1)H_{n-2}(x)-{1\over4}n(n-1)(n-2)(n-3)H_{n-4}(x).
\label{e403}
\end{equation}
The energy $E_n$ to order $\epsilon$ is
\begin{equation}
E_n=n+{1\over2}-{3\over4}\left( 2n^2 +2n + 1 \right)\epsilon+{\rm O}
(\epsilon^2).
\label{e404}
\end{equation}

We must also ${\cal PT}$ normalize the expression for $\phi_n(x)$ according to
(\ref{e7}) so that its square integral is $(-1)^n$:
\begin{equation}
\int_{-\infty}^{\infty} dx\,\left[\phi_n(x)\right]^2=(-1)^n+{\rm O}(\epsilon^2).
\label{e405}
\end{equation}
This determines the value of $a_n$ in (\ref{e402}). The result is very simple;
to order $\epsilon$ we have
\begin{equation}
a_n=1+{\rm O}(\epsilon^2).
\label{e406}
\end{equation}

Finally, we substitute $\phi_n(x)$ in (\ref{e402}) into (\ref{e14}) and use the
identity in (\ref{e37}). However, we obtain the trivial result that only the
leading term (zeroth-order in powers of $\epsilon$) survives. More generally,
we can show by a parity argument that the coefficients of {\em all} higher
powers of $\epsilon$ vanish. Thus, we get the (wrong) result that
\begin{equation}
{\cal C}(x,y)=\delta(x+y) \qquad{\rm (WRONG!)}.
\label{e407}
\end{equation}
We know that this result is wrong because the operator ${\cal C}(x,y)$ is
complex and the result in (\ref{e407}) is real. An alternative way to see this
is to note (\ref{e407}) imples that ${\cal C}(x,y)$ and ${\cal P}(x,y)$
coincide; but in this ${\cal PT}$-symmetric theory, ${\cal C}(x,y)$ and
${\cal P}(x,y)$ are distinct operators. We will see that the difference between
${\cal C}(x,y)$ and ${\cal P}(x,y)$ is a nonperturbative term of order
$e^{-1/(3\epsilon)}$, which is smaller than any integer power of $\epsilon$.

\subsection{Nonperturbative Analysis}
\label{ss4.2}

We will now show how to perform a nonperturbative analysis of the Schr\"odinger
equation (\ref{e401}). We decompose the eigenfunction $\phi_n(x)$ into its
perturbative part on the right side of (\ref{e402}) and a nonperturbative part:
\begin{equation}
\phi_n(x)= \phi_n^{\rm pert}(x)+ \phi_n^{\rm nonpert}(x).
\label{e408}
\end{equation}
The nonperturbative part of $\phi_n(x)$ is exponentially small compared with the
perturbative part, but these two contributions can be easily distinguished
because for real argument $x$, one is real while the other is imaginary.

Following the WKB analysis in Ref.~\cite{BMY}, we break the real-$x$ axis
into three regions: In region I, where $|x|\ll\epsilon^{-1/4}$, we have
\begin{eqnarray}
\phi_n^{\rm pert}(x)&\sim&{i^n\over\pi^{1/4}\sqrt{n!}}D_n(x\sqrt{2}),\nonumber\\
\phi_n^{\rm nonpert}(x)&\sim&ib_n C_n(x\sqrt{2}),
\label{e409}
\end{eqnarray}
where the coefficient of $D_n$ is taken from (\ref{e402}) and the coefficient
$ib_n$ of $C_n$ will be determined by asymptotic matching. Note that for
nonnegative integer index the parabolic cylinder function $D_n$ is expressed in
terms of a Hermite polynomial $H_n$ as
\begin{equation}
D_n(x\sqrt{2})=2^{-n/2}e^{-{1\over2}x^2}H_n(x).
\label{e410}
\end{equation}
Also, for nonnegative integer index the functions $D_n$ and $C_n$ are a pair
of linearly independent solutions to the parabolic cylinder equation. They
can be expressed in terms of parabolic cylinder functions as follows:
\begin{eqnarray}
D_n(z)&\equiv&{n!\over\sqrt{2\pi}}\left[i^n D_{-n-1}(iz)+(-i)^nD_{-n-1}(-iz)
\right],\nonumber\\
C_n(z)&\equiv&{i\over\sqrt{2\pi}}\left[i^n D_{-n-1}(iz)-(-i)^nD_{-n-1}(-iz)
\right].
\label{e411}
\end{eqnarray}

In region II, where $1\ll |x|\ll \epsilon^{-1/2}$, we can obtain the
eigenfunction using WKB theory. We write the Schr\"odinger equation (\ref{e401})
in the form $\phi_n''(x)=\omega_n(x)\phi_n(x)$ where, to leading order in
$\epsilon$, we have $\omega_n(x)=-2\epsilon x^4+x^2-2n-1$. Then, for positive
$x$ the physical-optics WKB approximation reads
\begin{eqnarray}
\phi_n^{\rm pert}(x)&\sim&f_n [\omega_n(x)]^{-1/4}\exp\left[-\int_{x_1}^x
ds\, \sqrt{\omega_n(s)}\right],\nonumber\\
\phi_n^{\rm nonpert}(x)&\sim&g_n [\omega_n(x)]^{-1/4}\exp\left[+\int_{x_1}^x
ds\, \sqrt{\omega_n(s)}\right],
\label{e412}
\end{eqnarray}
where the constants $f_n$ and $g_n$ will be determined by asymptotic matching.
The lower endpoint of integration, $x_1=\sqrt{2n+1}$, is the approximate
location of the inner turning point.

In region III $x$ is near the outer turning points at $\pm 1/\sqrt{2\epsilon}$.
For positive $x$ we define the variable $r$ by $x=x_2\left(1-2^{1/3}\epsilon^{2
/3}r\right)$, where $x_2=1/\sqrt{2\epsilon}$. The condition that $x$ is near
$x_2$ is that $r\ll\epsilon^{-2/3}$. In this region the Schr\"odinger equation
becomes an Airy equation in the variable $r$: $\phi_n''(r)=r\phi_n(r)$. The
solution in this region reads
\begin{eqnarray}
\phi_n^{\rm pert}(r)&\sim&h_n {\rm Bi}(r),\nonumber\\
\phi_n^{\rm nonpert}(r)&\sim&-i h_n {\rm Ai}(r),
\label{e413}
\end{eqnarray}
where ${\rm Ai}(r)$ and ${\rm Bi}(r)$ are the exponentially decaying and growing
Airy functions for large positive $r$. The fact that the same coefficient $h_n$
multiplies both ${\rm Bi}$ and ${\rm Ai}$ is a nontrivial result that is
established in Ref.~\cite{BMY}.

By asymptotically matching the solutions in regions I and II and the solutions
in regions II and III we obtain the formula for the coefficient of the
nonperturbative part of the solution in (\ref{e409}):
\begin{equation}
b_n=-{i^n\pi^{1/4}\over\sqrt{2\,n!}}(4/\epsilon)^{n+1/2}
e^{-{1\over3\epsilon}}.
\label{e414}
\end{equation}

Finally, using the wave function in region I we can construct the operator
${\cal C}(x,y)$ according to (\ref{e14}):
\begin{eqnarray}
{\cal C}(x,y) &=& \sum_{n=0}^\infty \phi_n(x) \phi_n(y)\nonumber\\
&=& \sum_{n=0}^\infty \left[
\phi_n^{\rm pert}(x)\phi_n^{\rm pert}(y)
+ \phi_n^{\rm pert}(x)\phi_n^{\rm nonpert}(y)
+ \phi_n^{\rm nonpert}(x)\phi_n^{\rm pert}(y)
+ \phi_n^{\rm nonpert}(x)\phi_n^{\rm nonpert}(y)\right].
\label{e415}
\end{eqnarray}
The first sum in this equation gives $\delta(x+y)$ to all orders in powers of
$\epsilon$ as explained above in Subsec.~\ref{ss4.1}. The last sum is negligible
compared with the second and third sums. We thus obtain
\begin{equation}
{\cal C}(x,y) = \delta(x+y)-i \sqrt{2/\epsilon}\,\, e^{-{1\over3\epsilon}}
\sum_{n=0}^\infty {1\over n!}(-4/\epsilon)^n
\left[D_n(x\sqrt{2}) C_n(y\sqrt{2})+ C_n(x\sqrt{2}) D_n(y\sqrt{2})\right],
\label{e416}
\end{equation}
where $C_n$ and $D_n$ are defined in (\ref{e411}). Observe that the correction
to the delta function (that is, the difference between the ${\cal P}$ operator
and the ${\cal C}$ operator) is nonperturbative; it is exponentially small and
imaginary.

The summation in (\ref{e416}) can be converted to a double integral:
\begin{equation}
{\cal C}(x,y)=\delta(x+y)+i\sqrt{2\over\pi^3\epsilon}\,\, e^{-{1\over3\epsilon}}
e^{{1\over2}(x^2+y^2)}\left\{{\partial\over\partial x}\int_0^\pi d\theta\int_0^1
{ds\over\sqrt{1+s^2}}\exp\left[{\left(2\sqrt{2s/\epsilon}\,\cos\theta-ix-isy
\right)^2\over1+s^2}\right] + (x\leftrightarrow y)\right\}.
\label{e417}
\end{equation}
This is the leading-order nonperturbative approximation to the coordinate-space
representation of the operator ${\cal C}$.

\vskip2pc This work was supported by the U.S.~Department of Energy.


\begin{references}

\bibitem{BB} C.~M.~Bender and S.~Boettcher, Phys.~Rev.~Lett.
{\bf 80}, 5243 (1998).

\bibitem{DDT} P.~Dorey, C.~Dunning and R.~Tatao, J.~Math.~Phys. A{\bf 34}, L391
(2001); {\em ibid}. A{\bf 34}, 5679 (2001). See also K.~C.~Shin, J. Math. Phys.
{\bf 42}, 2513 (2001) and Commun. Math. Phys. {\bf 229}, 543 (2002).

\bibitem{BBM} C.~M.~Bender and S.~Boettcher and P.~N.~Meisinger, J.~Math.~Phys.
{\bf 40}, 2201 (1999).

\bibitem{MAZM}
A.~Mostafazadeh, J.~Math.~Phys. {\bf 43}, 205 (2002); {\em ibid} {\bf 43}, 2814
(2002); {\em ibid} {\bf 43}, 3944 (2002); preprint (math-ph/0203041); preprint
(math-ph/0209018); Z.~Ahmed, Phys.~Lett. A{\bf 294}, 287 (2002);
G.~S.~Japaridze, J.~Phys.~A{\bf 35}, 1709
(2002); M.~Znojil, preprint (math-ph/0104012); A.~Ramirez and B.~Mielnik,
Working Paper 2002; C.~M.~Bender, D.~C.~Brody, L.~P.~Hughston, and
B.~K.~Meister, preprint (2002); D.~T.~Trinh, PhD Thesis, University of
Nice-Sophia Antipolis (2002), and references therein.

\bibitem{BBJ} C. M. Bender, D. C. Brody, and H. F. Jones, to be published in
Physical Review Letters.

\bibitem{BW} C.~M.~Bender and Q.~Wang, J.~Phys. A {\bf 34}, 3325 (2001);
C.~M.~Bender, S.~Boettcher, P.~N.~Meisinger, and Q.~Wang, Phys.~Lett.~A
{\bf 302}, 286 (2002).

\bibitem{BMY} C.~M.~Bender, P.~N.~Meisinger, and H. Yang, Phys.~Rev.~D
{\bf 63}, 045001-1 (2001).

\end{references}
\end{document}